# Identification of fast-changing signals by means of adaptive chaotic transformations

**Marek Berezowski, Marcin Lawnik**

Faculty of Applied Mathematics, Silesian University of Technology
Kaszubska str. 23, 44-100 Gliwice, Poland
marek.berezowski@polsl.pl; marcin.lawnik@polsl.pl

**Abstract**. The adaptive approach of strongly non-linear fast-changing signals identification is discussed. The approach is devised by adaptive sampling based on chaotic mapping "in yourself" of a signal. Presented sampling way may be utilized online in the automatic control of chemical reactor (throughout identification of concentrations and temperature oscillations in real-time), in medicine (throughout identification of ECG and EEG signals in real-time), etc. In this paper, we presented it to identify the Weierstrass function and ECG signal.

## 1 Introduction

One of fundamental problems in signals analyses and transformation is their identification, most often by means of sampling. The selection of an appropriate sampling method has been widely discussed in publications. The basic way of sampling was presented by Shannon [1], where the reconstruction of a signal is achieved on the basis of evenly collected samples. In accordance with Nyquist–Shannon theorem, the signal should be sampled with the frequency that is at least twice as big as the boundary frequency of its spectrum. However, the determination of the boundary frequency is not always possible in practice, especially in the case of fast-changing signals.

Under such circumstances sampling should be performed at uneven time intervals, for example: by adaptive sampling, in which the actual sampling moment depends on the previous sample value. Uneven and adaptive sampling methods were described in [2–15].

The scope of this paper is the adaptive sampling approach that is much easier than the so far described approaches. It is based on uneven adaptive sampling with the use of the chaotic representation of the tested signal. Unlike other popular methods, it does not require any additional analyses of the signal, such as, for example: Fourier's method, or determination of the frequency spectrum. As a result, the transformed signal is derived with good accuracy.

**Notations**: *a, b* –Weierstrass function parameters, *E* – information entropy, *N* – number of samples, observations horizon, *p* – probability, *t* – time, $\lambda$ – Lyapunov's exponent, *w(t)* – function generating nonlinear oscillations.

## 2 Sampling method

A fundamental problem encountered at identifying signal *w(t)* is an appropriate selection of the sampling rate. This is essential especially in the case of strongly non-linear fastchanging signals, for example: in a chemical reactor with recycle [16, 17] or in EEG signals [18], when even small time intervals between successive samples do not guarantee the detection of all signal changes.

Such inconvenience may be evaded by using the mapping "in yourself" of the signal involving the adaptive sampling at time intervals $t_k$, designating the values of the samples, in accordance with the following recursive equation:

$$t_{k+1} = w(t_k). \tag{1}$$

The condition for the use of this method is that the sequence generated by Eq. (1) is chaotic. Then the derived information about the investigated signal is the best, as confirmed by the information entropy of the information of sequence (1)

$$E = -\sum_{k=1}^{N} p_k \log_2 p_k = \log_2 N \tag{2}$$

where: *N* is the observation horizon, whereas $p_k$ is the probability of designating moment $t_k$, at which the sample should be collected. Due to the of uniqueness of the elements of sequence (1) particular probabilities are the same, equal to $p_i = 1/N$. For the infinitely long observation horizon, the entropy has an infinite value, which may be directly inferred from equation (2). Accordingly, a complete reconstruction of function *w(t)* is derived.

## 3 Example $\omega(t)$ as Weierstrass function

A good example for the application of the discussed approach is the identification of the signal, where the right-hand side of Eq. (1) is the Weierstrass function [19–21]

$$w(t) = \sum_{i=1}^{\infty} a^i \cos(\pi b^i t) \tag{3}$$

where 0 < *a* < 1, whereas *b* is an odd number fulfilling condition $ab > 1 + 1.5\pi$, as the signal is infinitely fast-changing, which means that there is a significant change of its value at each moment of time. In consequence, the derivative of the signal in relation to time: $d\omega/dt$ has an infinitely big value at each time moment *t*. Thus, under such circumstances it is not possible to select a sampling range that would guarantee the accurate identification of the function. An exemplary graph of function *w(t)* is shown in Fig. 1.

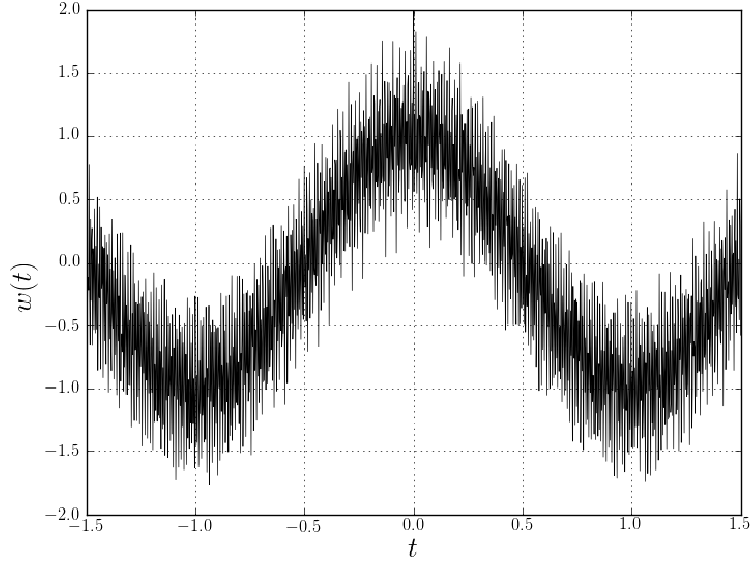

Fig. 1. Graph of the Weierstrass function generated from equation (3). $a = 0.5$, $b = 571$.

In accordance with (1), the recursive transformation for the Weierstrass function (3) assumes the following form:

$$t_{k+1} = \sum_{i=1}^{\infty} a^i \cos(\pi b^i t_k). \qquad (4)$$

As shown in the above analysis, the identification of signal $w(t)$ is relevant only if Eq. (4) generates a chaotic sequence of the elements $t_k$. Thus, a possibility of the occurrence of such sequence should be checked, for example, by determining the sign of Lyapunov's exponent concerning transformation (4). In view of the nature of the Weierstrass function derivative, Lyapunov's exponent

$$\lambda = \lim_{n \to \infty} \frac{1}{n} \sum_{k=1}^{n-1} \ln \left| \frac{dw(t)}{dt} \right|_{t=t_k} \qquad (5)$$

has always an infinitely big positive value. Accordingly, the sequence generated by transformation (4) is chaotic. Exemplary distribution of the samples of function $w(t_k)$ derived from transformation (4) is shown in Fig. 2, where the successive samples are marked on the horizontal axis. Figure 2 also reflects the phenomenon of infinite sensitivity of the recursive process to the change of the initial conditions. Such great sensitivity results from an infinitely big value of Lyapunov's exponent (5). In Fig. 2, variables $t_0$ differ at moment $k = 0$ with the value equal only to $10^{-16}$. Despite such small difference at the beginning, the values the variables are significantly different even after the first step. The nature of Lyapunov's exponent for 100 elements the Weierstrass sequence is shown in Fig. 3. It should also be noted that although the solutions of Eq. (4) are chaotic, the graph of Lyapunov's exponent as in Fig. 3 does not have a fractal nature.

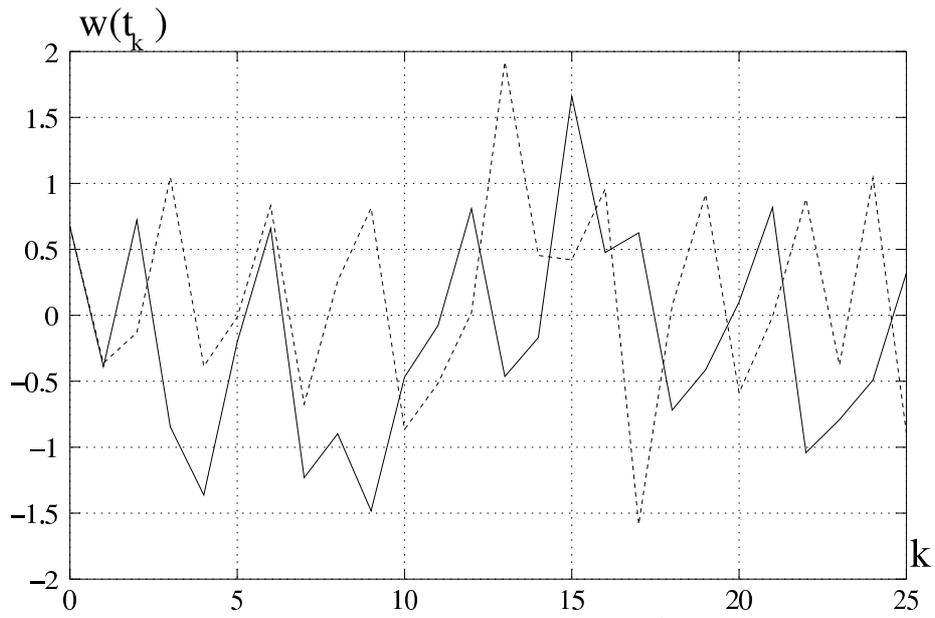

Fig. 2: Sampling distribution. $a = 0.5$, $b = 571$.

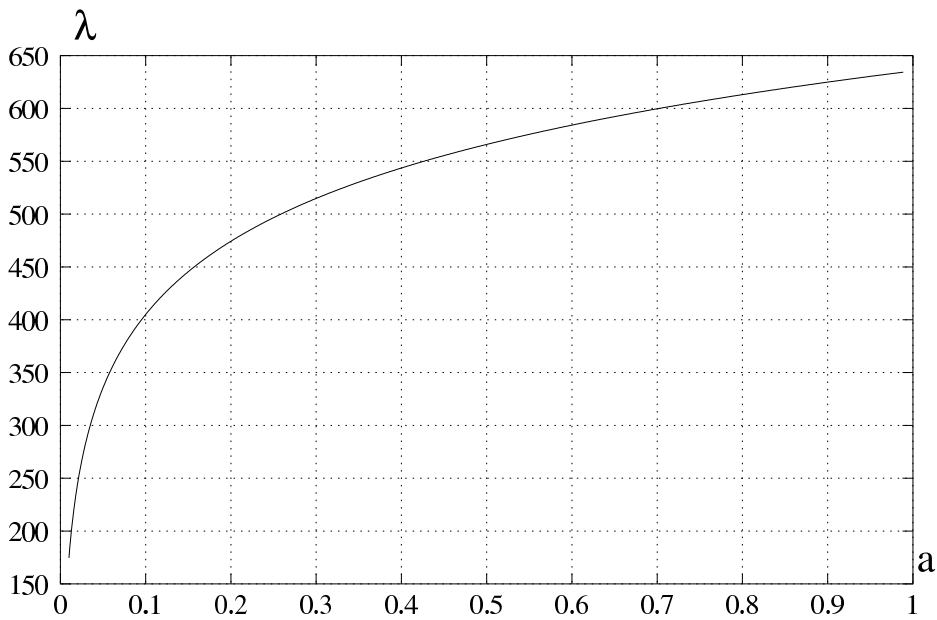

Fig. 3: Graph of Lyapunov's exponent. $b = 571$.

To illustrate the discussed identification method, exemplary detection of signal *w(t)* for *a*=0.5 was performed – see Fig. 4, which is very consistent with the graph derived from Eq. (1) (Fig. 1).

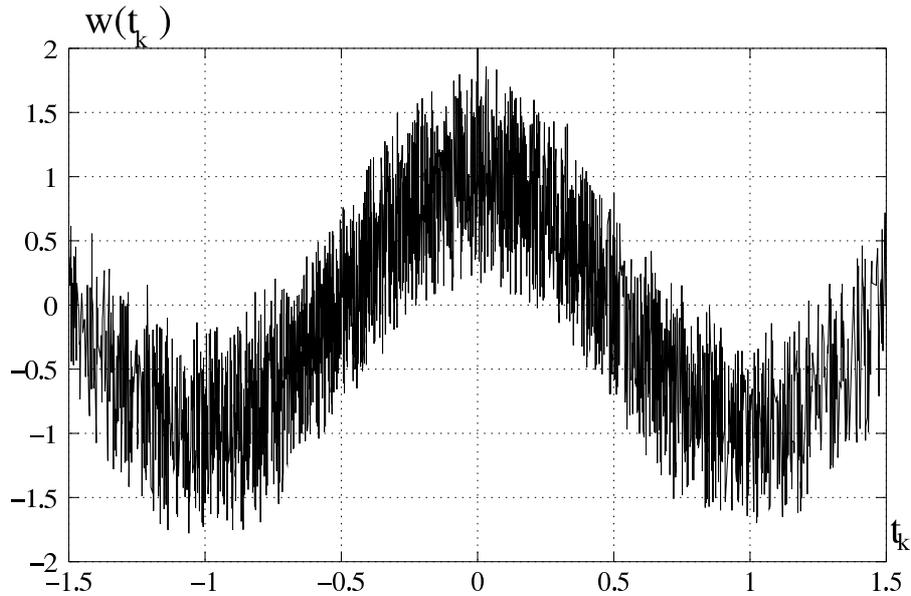

Fig. 4: Graph of the Weierstrass function generated from equation (4). $a=0.5, b=571$.

It should also be stressed that the graph in Fig. 4 is, at the same time, a strange attractor of transformation (4).

## 4  Example w(t) as ECG signal

Presented in this paper sampling approach may be also utilized in medicine for identification of ECG signals in real-time. It can be used in particular to accurately determine the actual amplitude and frequency of the measured signals. This allows the doctor or the automatic control device to react in time and in the right way. Further analysis we have relied on sample ECG results available in [22] (thin line in Fig. 5). Therefore the right-hand side of Eq. (1) is not now a continuous function as before, but a discrete set of measurement results. The part of the ECG signal was reconstructed using an identification procedure (1) (large dots in Fig. 5). In practice, the method is best for a on-line continuous signal.

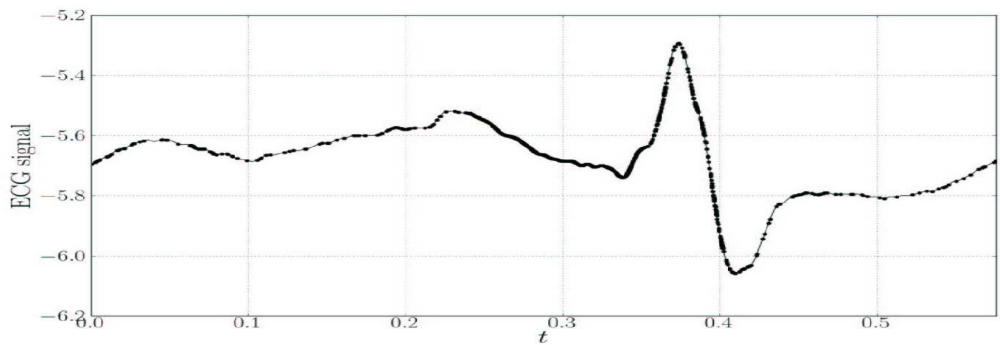

Fig. 5. ECG signal identification. Thin line – actual ECG signal; dots – identification result.

## 5  Concluding remarks

The discussed method of signal identification by means of uneven adaptive sampling based on recursive chaotic transformation may be applied, first and foremost, for the identification of strongly non-linear fast-changing signals. An exemplary use of the methodwas shown for the identification of the signal generated by the Weierstrass function. The method may also be utilized in the automatic control process based on the observation of the fast-changing signal in real-time. Unlike other popular approaches, it does not require additional analyses of the identified signal, or Fourier's analysis, or determination of the frequency spectrum. The method is easy in practical application and renders accurate representation of the original signal.